\documentclass[prb,twocolumn,amsmath,showpacs]{revtex4}

\usepackage{graphicx}

\begin{document}
\input epsf

\title {Temperature dependence of the upper critical field of type-II superconductors 
from isothermal magnetization data. Application to high temperature superconductors}

\author {I. L. Landau$^{1,2}$ H. R. Ott$^{1}$}
\affiliation{$^{1}$Laboratorium f\"ur Festk\"orperphysik, ETH H\"onggerberg, 
CH-8093 Z\"urich, Switzerland}
\affiliation{$^{2}$Kapitza Institute for Physical Problems, 117334 Moscow, 
Russia}

\date{\today}

\begin{abstract}
Using the Ginzburg-Landau theory in very general terms, we develop a simple scaling 
procedure which allows to establish the temperature dependence of the upper critical 
field $H_{c2}$ and the value of the superconducting critical temperature $T_{c}$ of 
type-II superconductors from measurements of the reversible isothermal magnetization. 
An analysis of existing experimental data shows that the normalized dependencies of 
$H_{c2}$ on $T/T_{c}$ are practically identical for all families of high-$T_{c}$ 
superconductors at all temperatures for which the magnetization data are available.
\end{abstract}

\maketitle

\section {Introduction}

Establishing the upper critical field $H_{c2}$ and its temperature dependence from 
experimental data is not a simple task for high-$T_{c}$ superconductors 
(HTSC\textquoteright s). The 
main problem is that the transition to the superconducting state, probed by either 
measuring the magnetization $M(T)$ of the sample or its resistance $R(T)$, does not 
reveal any sharp features in $M(T)$ or $R(T)$ around $H_{c2}$. The notorious widths 
of the transitions are usually attributed to fluctuation effects. One of the ways to 
overcome this difficulty is to use the temperature dependence of the magnetic moment 
of the sample $M(T)$, measured in a constant external magnetic field $H$, and to 
extrapolate the linear part of the $M(T)$ curve to the $M$ value corresponding to 
the normal state of the sample. \cite{1}$^-$\cite{5} This procedure is usually 
justified by invoking the Abrikosov theory of the mixed state according to which 
the magnetization per unit volume may by written as 
\begin{equation}
M(H,T)={1 \over {4\pi }}{{H_{c2}(T)-H} \over {(2\kappa ^2-1)\beta _A}},
\end{equation}
were $\kappa$ is the Ginzburg-Landau (GL) parameter and $\beta _{A}= 1.16$ for a 
triangular vortex lattice. \cite{6,7} From Eq. (1) it follows naturally that the 
magnetic moment varies linearly with temperature for a fixed value of the external 
field, if $H_{c2}(T)$ is a linear function of temperature. The problem is that Eq. 
(1) is only valid close to $H_{c2}$. Nevertheless, most experimental M(T) curves 
are practically linear in $T$ for magnetic fields between $0.1H_{c2}$ and 
$0.6H_{c2}$, \cite{1} i.e., well below $H_{c2}$. In this range of magnetic fields, 
the magnetic moment is certainly a non-linear function of $[H_{c2}(T) - H]$. The 
apparent linearity of the experimental $M(T)$ curves is most likely the result of 
some non-linearity of $H_{c2}(T)$. In this situation, a simple linear extrapolation 
of $M(T)$ curves will almost certainly result in wrong $H_{c2}(T)$ curves. As an 
example of this type of failure, we mention a study of 
Bi$_{2}$Sr$_{2}$CaCu$_{2}$O$_{8}$ single crystals where the application of this 
extrapolation procedure resulted in completely unphysical $H_{c2}(T)$ curves. 
\cite{8}

Another method for establishing $H_{c2}(T)$ is to use theoretical calculations of 
$M(H)$ extended to $H << H_{c2}$ in order to evaluate $H_{c2}$ from experimental 
$M(H)$ or $M(T)$ curves. This approach is again not very reliable. First of all, 
solving the GL equations in two dimensions for magnetic fields well below $H_{c2}$ 
represents a formidable mathematical task. To our knowledge, there is only one study 
in which this problem has been solved numerically for the Abrikosov vortex lattice 
for a selected set of values of the GL parameter $\kappa$. \cite{9} However, as far 
as we are aware, nobody has tried to use the results of Ref. 9 for the 
interpretation of the experimental data. More often, approximate models for the space 
dependence of the order parameter in the vortex structures are used. The most popular 
is the Hao-Clem model,\cite{10} which has widely been used to derive different 
parameters of HTSC\textquoteright s from magnetization data. \cite{11}$^-$\cite{27} 
However, as has been pointed out recently, this 
model is far from being accurate. \cite{28,29} Both the Hao-Clem model and the 
numerical calculations in Ref. 9 assume uniform and isotropic superconductors, 
i.e., conditions that are definitely not met in the case of HTSC compounds. 

Although the dependence of the sample resistance on temperature, $R(T)$, in external 
magnetic fields is often used for the evaluation of $H_{c2}(T)$, 
\cite{30}$^-$\cite{40} we believe that this approach is even less reliable than 
the use of magnetization measurements. The transition to the normal state resistance 
is very gradual and there is no appropriate theory for an interpretation of $R(T)$ 
curves. It is quite likely that the misinterpretation of the resistance data is the 
main reason why $H_{c2}(T)$ curves derived from the results of resistance measurements 
often exhibit an unusual positive curvature.

In order to evaluate $H_{c2}$ from the experimental data in such complicated materials 
as HTSC\textquoteright s, it is very important to introduce an appropriate definition 
of the upper critical field. In an ideal type-II superconductor, $H_{c2}$ is the 
highest value of a magnetic field compatible with superconductivity, i.e., the 
$H_{c2}(T)$ curve on the $H - T$ phase diagram represents a line of second order phase 
transitions to the normal state. As is well known for HTSC superconductors, this 
transition degenerates to a cross-over region because of fluctuation effects and even 
in magnetic fields $H > H_{c2}(T)$ superconducting features appear in the data of 
resistivity and magnetization measurements. We also note that small inclusions of 
another superconducting phase with a higher or lower critical temperature, $T'_c$, than 
that of the bulk cannot always be excluded in HTSC\textquoteright s. In magnetic fields 
$H > H_{c2}$ the impact of such inclusions with $T'_c>T_{c}$ on the sample resistance 
or its magnetization is similar to that arising from superconducting fluctuations. At 
the same time, in magnetic fields well below $H_{c2}(T)$, the effect of fluctuations 
and possible inclusions of impurity phases on the sample magnetization is small and 
the $M(H)$ curves in this magnetic field range must be practically the same as for the 
perfectly uniform sample without fluctuations. This circumstance provides the 
possibility to evaluate the temperature dependence of $H_{c2}$, in its traditional 
sense, from such magnetization measurements.

In this paper we propose a new approach to this problem by scaling the $M(H)$ curves 
measured at different temperatures. This scaling procedure is based on the application 
of the GL theory, without assuming any specific magnetic field dependence of the 
magnetization. In this way one can only establish the temperature dependence of 
$H_{c2}$, but its absolute values remain unknown. Below we describe the method in 
detail and apply it to experimental data available in the literature. It turns out 
that in many cases the extrapolation of the normalized $H_{c2}(T)$ curve to 
$H_{c2} = 0$ provides reliable values of the superconducting critical temperature 
$T_{c}$. 

\section {Scaling procedure}

Our scaling procedure is based on the assumption that the Ginzburg-Landau (GL) 
parameter $\kappa$ is temperature independent. Although the microscopic theory of 
superconductivity predicts a temperature dependence of $\kappa$, \cite{41,42} this 
dependence is rather weak and is not expected to change the results significantly. 
From the GL theory it follows straightforwardly that, if $\kappa$ is temperature 
independent, the magnetic susceptibility $\chi$ of the sample is a universal function 
of $H/H_{c2}$, i.e., $\chi (H,T) = \chi (h)$ with $h = H/H_{c2}(T)$, \cite{6}. The 
magnetization density is
\begin{equation}
M(H,T)=H_{c2}(T)h\chi (h).
\end{equation}
According to Eq. (2) the sample magnetization, for the same value of $h = H/H_{c2}$, 
is proportional to $H_{c2}(T)$. This leads to the following relation between the 
values of $M$ at two different temperatures, $T_{0}$ and $T$, 
\begin{equation}
M(H,T_0)=M(h_{c2}H,T)/h_{c2}
\end{equation}
with $h_{c2} = H_{c2}(T)/H_{c2}(T_{0})$. The collapse of individual $M(H)$ curves 
measured at different temperatures may be achieved by a suitable choice of $h_{c2}(T)$. 
Of course, the scaling procedure implied by Eq. (3) is valid for ideal type-II 
superconductors only and in the following we consider the necessary corrections to Eq. 
(3) that are dictated by some specific features of HTSC\textquoteright s. 

Most of the families of HTSC\textquoteright s reveal a weak paramagnetic susceptibility 
$\chi _{n}$ in the normal state. \cite{5,13}$^-$\cite{19,43}$^-$\cite{46} Its 
influence may be accounted for by replacing Eq. (3) by
\begin{equation}
M(H,T_0)=M(h_{c2}H,T)/h_{c2}+c_0(T)H,
\end{equation}
where $c_{0}(T)= \chi_{n}(T_{0}) - \chi_{n}(T)$.

For many HTSC materials the derivative $dM/dH$ changes its sign when approaching the 
critical temperature from below. \cite{8,14}$^-$ \cite{22,43}$^-$\cite{45,47}$^-$\cite{49} 
Because the field dependence of the magnetization in the mixed state always requires 
$dM/dH > 0$, \cite{6} the change of sign of $dM/dH$ cannot be explained by 
considering the properties of a static mixed state alone. This sign change is usually 
attributed to fluctuation effects. We assume that the additional contributions to the 
magnetization arising from fluctuation effects may be described by an effective 
susceptibility $\chi_{eff}(T)$, which is independent of the applied magnetic field. 
In this case we can still use Eq. (4) but with $c_{0}(T) = [\chi_{n}(T_{0} - 
\chi_{n}(T)] + [\chi_{eff}(T_{0}¥) - \chi_{eff}(T)]$. In the following we use the 
parameter $c_{0}¥(T)$ in Eq. (4) as an additional adjustable parameter in the scaling 
procedure. The assumption that $\chi_{eff}(T)$ does not depend on the magnetic field 
is a simplification and this is why Eq. (4) should not be used in the temperature 
range where $dM/dH < 0$ and where the fluctuation-induced magnetization dominates the 
magnetic moment of the sample.  

We note that the term $c_{0}H$ in Eq. (4) may also account for any contribution to 
the magnetization arising from small inclusions of another superconducting phase with 
a different $T_{c}$. If the value of $T_{c}$ of this minority phase is higher than 
that for the bulk of the sample, some small regions of the sample will remain in the 
superconducting state even if $H > H_{c2}(T)$. These superconducting islands also 
give a non-zero magnetic moment with $dM/dH < 0$. In magnetic fields $H < H_{c2}$, 
the contribution from these regions, where superconductivity is stronger than in the 
bulk of the sample, is superimposed onto the contribution to the magnetic moment 
arising from the mixed state. 

At this point, we wish to comment on the physical relevance of $H_{c2}(T)$ and $T_{c}$ 
obtained in this way. Because our analysis is based on measurements of the 
magnetization in the mixed state, $H_{c2}(T)$ corresponds to the disappearance of the 
mixed state rather than to a complete suppression of superconductivity in the sample. 
In magnetic fields $H > H_{c2}$ superconducting regions may, as mentioned above, still 
exist in the form of separated islands formed due to either thermal fluctuations or 
inclusions of minority phases with enhanced values of $T_{c}$ and $H_{c2}$. Similar 
superconducting islands may also exist in zero magnetic field at $T > T_{c}$. It is 
only important that the lateral extension of these islands is small enough, such that 
no mixed state can be established inside the island. The cause of such superconducting 
islands is not important for our consideration. It may be due to thermal fluctuations 
or sample inhomogeneities, as well as a combination of both. Note that for an ideal 
type-II superconductor without fluctuations these definitions of $H_{c2}(T)$ and 
$T_{c}$ coincide with the values of magnetic fields and temperature fixed be the onset 
of superconductivity. 

\section {Analysis of experimental data}

\begin{figure}[b]
 \begin{center}
  \epsfxsize=0.9\columnwidth \epsfbox {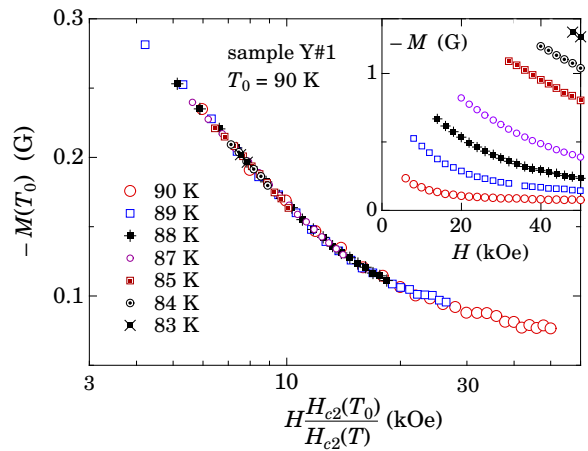}
  \caption{The magnetization data for sample Y\# 1 after scaling using Eq. (4) 
           with $T_{0} =$ 90 K. The inset displays the original data. \cite{46}}
 \end{center}
\end{figure}
We now apply our scaling procedure to experimental results available in the 
literature. As it turns out, the relative temperature variations of $H_{c2}$ are 
identical for practically all HTSC materials. Because this is a completely unexpected 
and, in our view, rather important result, we describe the analysis in some detail. 
We have analyzed magnetization data for 29 samples presented in 25 publications. Some 
information concerning these samples is listed in Tables I to IV. Letters in the 
sample identification denote the chemical element characterizing the considered 
family of HTSC\textquoteright s. Because the sample homogeneity is important for the 
applicability of our method, only single crystals and grain-aligned samples have been 
chosen. We have also limited our analysis to studies in which the magnetization 
measurements were extended up to temperatures $T \ge 0.94-0.95 T_{c}$, because only 
in these cases we may expect a reliable evaluation of $T_{c}$ by extrapolating 
$H_{c2}(T)$ to $H_{c2} = 0$.

\begin{figure}[!t]
 \begin{center}
  \epsfxsize=0.9\columnwidth \epsfbox {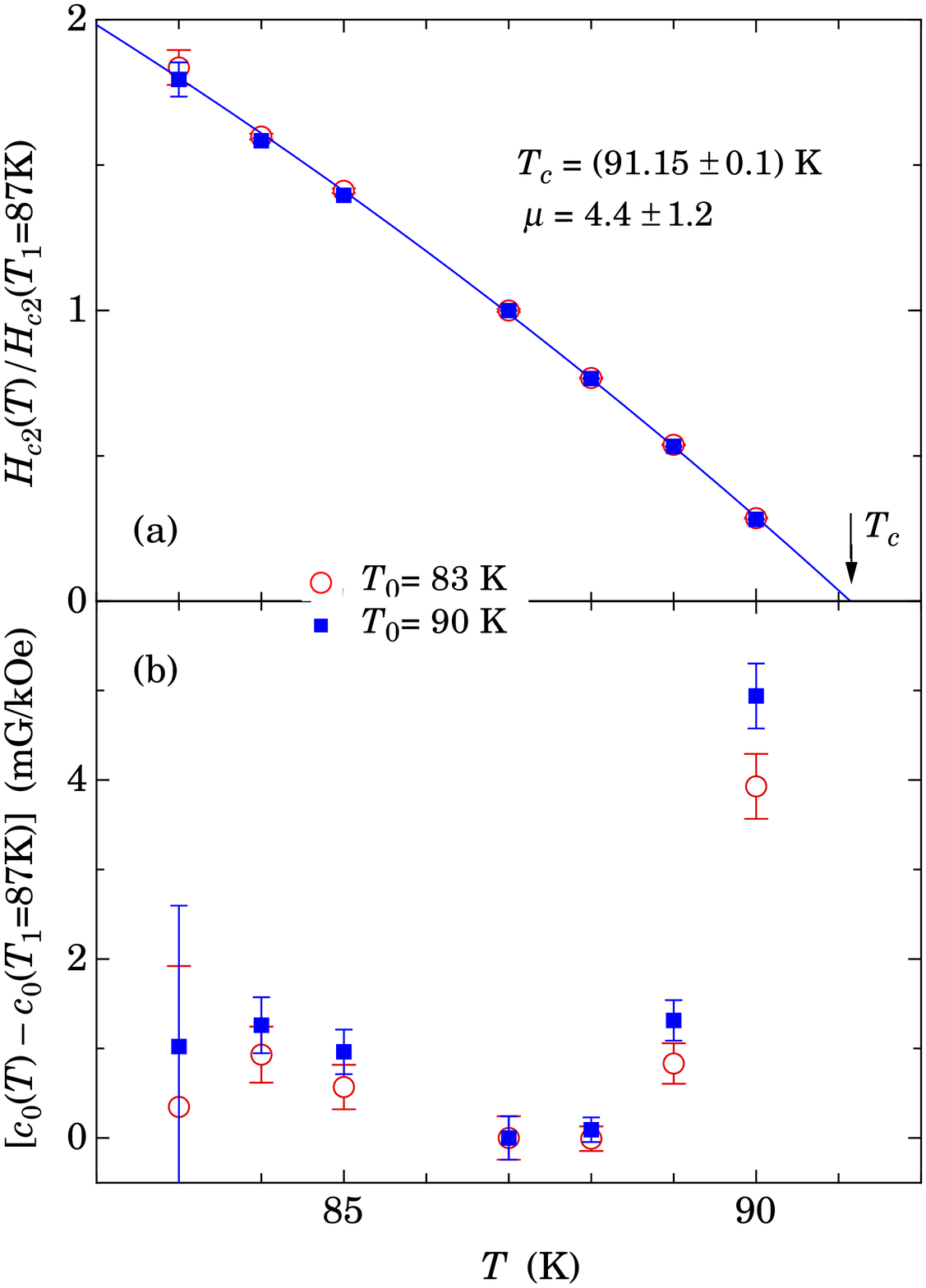}
  \caption{(a) $H_{c2}(T)/H_{c2}(87K)$ and (b) $[c_{0}(T) - c_{0}(87K)]$ resulting 
           for two different choices of $T_{0}$. The solid line is the best fit with 
		   Eq. (6).}
 \end{center}
\end{figure}
In order to make use of Eq. (4), the following procedure was employed. First, the 
$M(H)$ curve for some temperature $T = T_{0}$ was approximated by
\begin{equation}
M(H)=h_{c2}\left\{ {\sum\limits_{i=0}^n {A_i[\ln (H/h_{c2})]^i+c_0H}} \right\}
\end{equation}
with $h_{c2} = 1$ and $c_0 = 0$. The coefficients $A_{i}$ were used as fit parameters 
and the number $n$ was chosen such that a further enhancement of its value had no 
influence on the deviation parameter $\sigma$ of the approximation of $M(H)$. 
\cite{50} In the next steps, the coefficients $A_{i}$ were fixed and the parameters 
$h_{c2}$ and $c_{0}$ were evaluated via the fitting procedure for approximating the 
available $M(H)$ curves measured at other temperatures $T = T_{i}$. \cite{51} The 
result of this scaling procedure, representing the field dependence of the 
magnetization of sample Y\#1 at $T = T_{0}$, is shown in Fig. 1. It may be seen that 
a rather perfect overlap of the individual $M(H)$ curves measured at different 
temperatures, which are displayed in the inset of Fig. 1, is obtained in this way. 
Because the renormalized field variable $h_{c2}$ enters the denominator of Eq. (4), 
the magnetization data sets for the highest temperatures are considerably expanded 
along the vertical axis in comparison with low temperature data,. This is the reason 
for the somewhat enhanced scatter in the high temperature data.
\begin{figure*}[t]
 \begin{center}
  \epsfxsize=1.9\columnwidth \epsfbox {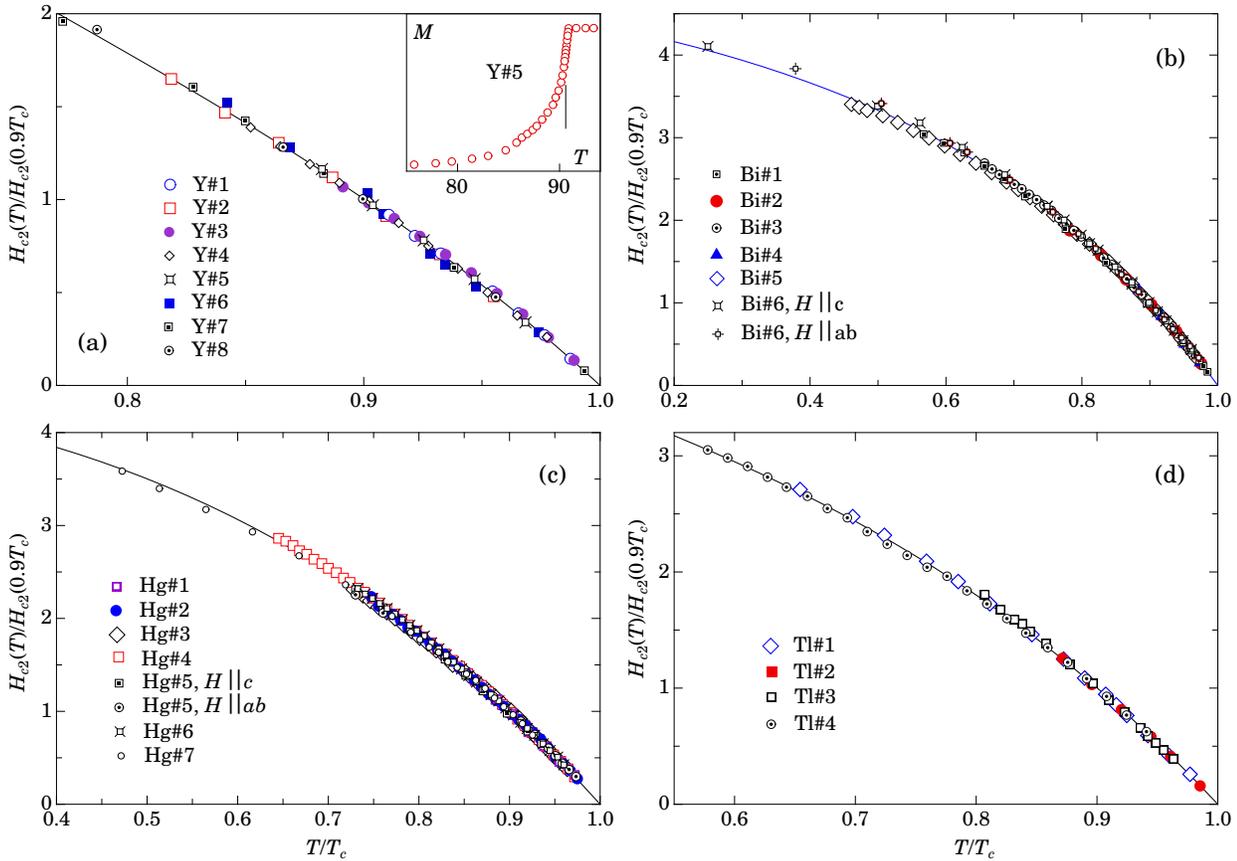}
  \caption{$H_{c2}(T)/H_{c2}(0.9T_{c})$ versus $T/T_{c}$ for different samples. The 
           solid lines are guides to the eye. (a) Y-based samples (Table I). The 
		   inset shows the low field $M(T)$ curve for sample Y\# 5. The vertical line 
		   in the inset indicates the value of $T_{c}$ as obtained by extrapolation 
		   of the corresponding $h_{c2}(T)$ curve.  (b) Bi-based samples (Table II). 
		   (c) Hg-based samples (Table III). (d) Tl-based samples (Table IV)}
 \end{center}
\end{figure*}

In order to demonstrate the consistency of our procedure, we show typical data sets 
for $H_{c2}(T)$ and $c_{0}(T)$ in Figs. 2(a) and 2(b), respectively. In this 
particular case, the scaling procedure was done twice, with $T_{0}$ either at the 
upper or at the lower end of the covered temperature range. In order to compare the 
results obtained in these two cases, $H_{c2}(T)$ and $c_{0}(T)$ are normalized by 
their values at $T = T_{1} = 87$ K. As may be seen, the result is practically 
independent of the choice of $T_{0}$. The parameter $c_{0}(T)$ in Eq. (4) accounts 
for only a small correction to $M(H)$. This causes a much enhanced uncertainty in the 
values of $c_{0}(T)$ than that for the normalized upper critical field, as may easily 
be seen by comparing Figs. 2(a) and 2(b).

We note that the uncertainty of $h_{c2}(T)$ increases considerably for temperatures 
close to $T_{c}$ as well as for the lowest temperatures. The loss of accuracy for 
the highest temperatures is due to the obvious enhancement of the experimental 
uncertainty of the $M(H)$ data. Although the accuracy is improving with decreasing 
temperature, the increase of the irreversibility field limits the available magnetic 
field range as may clearly be seen in the inset to Fig. 1. If the experimental data 
are collected in a narrow magnetic field range only, our scaling procedure is not 
reliable.

The temperature dependence of the normalized upper critical field, as shown in Fig. 
2(a), may also be used to evaluate the critical temperature $T_{c}$. For this purpose 
the ratio $H_{c2}(T)/H_{c2}(T_{1})$ was approximated by 
\begin{equation}
{{H_{c2}(T)} \over {H_{c2}(T_1)}}={{1-(T/T_c)^\mu } \over {1-(T_1/T_c)^\mu }},
\end{equation}
in which $\mu$ and $T_{c}$ are used as fit parameters. Eq. (6) provides a rather 
good approximation to $h_{c2}(T)$ curves for $T \ge 0.8T_{c}$. The corresponding fit 
is shown as the solid line in Fig. 2(a). The values of $\mu$ and $T_{c}$ are 
indicated in Fig. 2. \cite{52} If the experimental data were obtained up to 
temperatures rather close to the critical temperature, the extrapolated value of 
$T_{c}$ is quite accurate. A reliable value of $T_{c}$ is essential for the comparison 
of the results that were obtained for the samples with different critical temperatures. 
Using the values of $T_{c}$ evaluated in such a way, we have plotted 
$H_{c2}(T)/H_{c2}(0.9T_{c})$ versus $T/T_{c}$ as shown in Fig. 3(a). Quite 
surprisingly, the temperature variations of $H_{c2}$ for different Y-based compounds 
and different types of samples turn out to be identical. In the inset of Fig. 3(a) we 
display the low field magnetization curve $M(T)$ of sample Y$\# $5 and indicate the 
position of $T_{c}$ resulting from our extrapolation procedure with a vertical line. 

The temperature variations of $h_{c2}$ for other families of HTSC\textquoteright s are 
plotted in Figs. 3(b) - 3(d). Similar to what has been found for Y-based compounds, the 
scaling procedure again leads to an almost perfect merging of all the data onto one 
single curve for different samples. Furthermore, as may clearly be seen in Fig. 4, the 
temperature dependencies of the normalized upper critical field for different families 
of HTSC\textquoteright s are practically identical at all temperatures for which the 
experimental data are available. We note that the insignificant differences between the 
$h_{c2}(T/T_{c})$ curves for different samples, visible at the lowest temperatures in 
Figs. 3(a), 3(b), and 4, are due to small errors in the determination of the critical 
temperature. For the data presented in Figs. 3(a-d) the relative errors in the 
determination of the critical temperature, $\Delta T_{c}/T_{c}$, are between 0.001 and 
0.003, depending on the quality of the original experimental data. Although this 
uncertainty is quite small, it is sufficient to explain the observed differences 
between the $h_{c2}$ values at low temperatures.

\begin{figure}[t]
 \begin{center}
  \epsfxsize=0.9\columnwidth \epsfbox {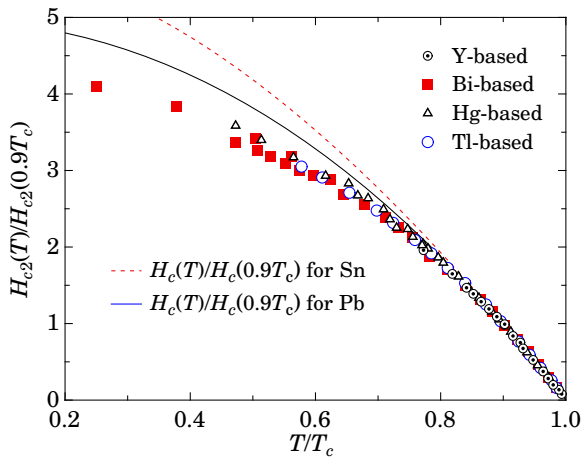}
  \caption{The normalized temperature dependence of $H_{c2}$ for different HTSC 
           compounds. The solid and broken lines represent the ratios 
		   $H_c(T)/H_c(0.9T_c)$ for pure metallic Lead and Tin, respectively.}
 \end{center}
\end{figure}
\begin{figure}[h]
 \begin{center}
  \epsfxsize=0.9\columnwidth \epsfbox {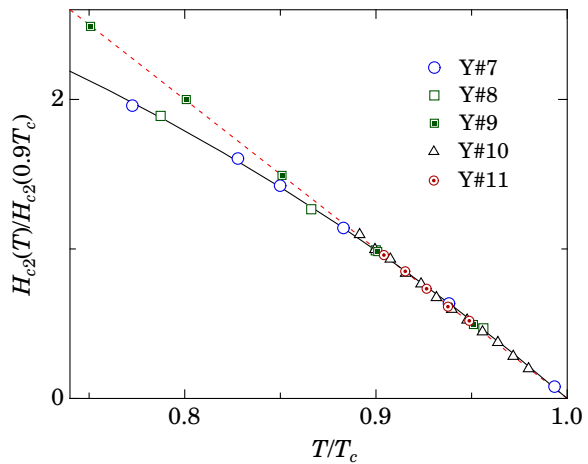}
  \caption{$H_{c2}(T)/H_{c2}(0.9T_c)$ data for several YBa$_2$Cu$_3$O$_{7Ðx}$ samples 
           with different oxygen contents. The solid line represents 
		   $H_{c2}(T)/H_{c2}(0.9T_c)$ for the Y-based samples shown in Fig. 3(a). The 
		   dashed line is the best fit to the data for sample Y\# 9, assuming a linear 
		   temperature variation of $H_{c2}$.}
 \end{center}
\end{figure}
Among the numerous samples listed in Tables I - IV, only for the oxygen deficient 
sample Y$\# $9, the $h_{c2}(T)$ curve is distinctly different.  As may be seen in 
Fig 5, $h_{c2}(T)$ for this sample is perfectly linear in the entire covered 
temperature range, in striking difference to two other, over-doped and optimally doped 
Y-based samples (Y\#7 and Y\#8) investigated in the same study. The magnetization data 
for sample Y\#9 were collected in a very wide range of magnetic fields and, as may be 
seen in Fig. 6, our scaling leads again to a nearly perfect merging of the curves. The 
difference in $h_{c2}(T)$ between Y\#9 and other samples is thus not due to 
insufficient sample quality but rather reflects the intrinsic difference in properties 
of under-doped YBa$_{2}$Cu$_{3}$O$_{7-x}$ materials. Only very few magnetization 
studies of oxygen deficient YBa$_{2}$Cu$_{3}$O$_{7-x}$ single crystals or grain-aligned 
samples are available in the literature and we could find only two additional 
publications which are suitable for our analysis (samples Y\#10 and Y\#11). \cite{5,53} 
Unfortunately, as may be seen in Fig. 5, the measurements reported in Refs. 5 and 53 
were made at temperatures very close to $T_{c}$ and in this temperature range, the 
temperature dependence of $H_{c2}$ is linear for all HTSC materials.

\begin{figure}[h]
 \begin{center}
  \epsfxsize=0.9\columnwidth \epsfbox {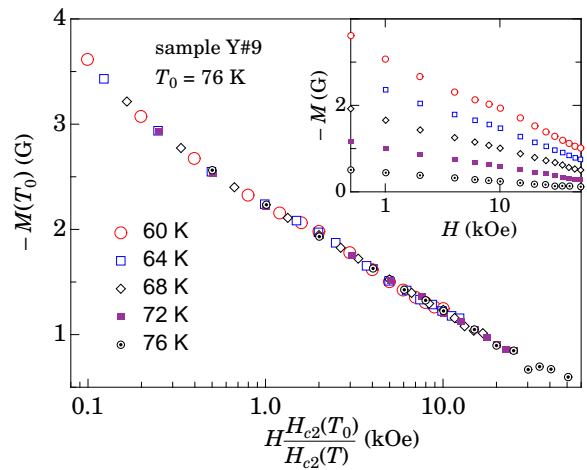}
  \caption{The magnetization data for sample Y\# 9 after scaling using Eq. (4) 
           with $T_0 =$ 76 K. The inset displays the original data.\cite{11}}
 \end{center}
\end{figure}
HTSC\textquoteright s are strongly anisotropic and it is well known that, depending on 
the orientation of the applied magnetic field, the absolute values of $H_{c2}$ differ 
significantly. This is why it is interesting to compare the results of our analysis 
for different orientations of the magnetic field. Unfortunately, we have found only 
two data sets from magnetization measurements that were made on the same samples but 
with two different orientations of the external magnetic field (Hg$\#$5 and Bi$\#$6). 
As may be seen in Fig. 3(c), the results for the grain-aligned sample Hg$\#$5 are 
practically independent of the orientation of the magnetic field. The situation for 
the single-crystal sample Bi$\# $6 is different. The resulting $h_{c2}(T)$ curves for 
this sample are shown in Fig. 7. The data perfectly match each other if we assume that 
the value of $T_{c}$ depends on the orientation of the magnetic field. This at first 
glance rather strange result may easily be understood if we recall our definition of 
$H_{c2}$ at the end of Section 2 and we discuss this point in the next paragraph. 

\begin{figure}[t]
 \begin{center}
  \epsfxsize=0.9\columnwidth \epsfbox {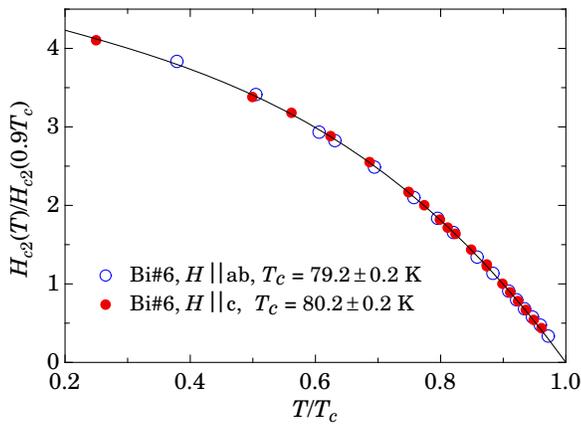}
  \caption{$H_{c2}(T)/H_{c2}(0.9T_{c})$ data for sample Bi\# 8 for 2 different 
           orientations of the external magnetic field.}
 \end{center}
\end{figure}
As has been described above, our procedure provides $H_{c2}$ as it enters the equation 
for the magnetization in the mixed state, i.e., the resulting $H_{c2}(T)$ curve in the 
$H$ - $T$ phase diagram represents the upper boundary for the existence of the mixed 
state. In a perfect type-II superconductor without fluctuations this definition of 
$H_{c2}$ coincides with the upper limit for the existence of superconductivity. In real 
HTSC materials the situation is different and superconducting regions may still exist 
in the sample even above the $H_{c2}(T)$ curve due to, for instance, thermal 
fluctuations or impurities with a higher $T_{c}$. We consider it as an advantage that 
such effects have practically no influence on our evaluation of $H_{c2}$. The situation 
is further complicated by the layered structure of HTSC\textquoteright s. As has 
previously been established by resistance measurements in zero magnetic field for 
Bi-based compounds, the superconducting coherence in the Cu-O planes sets in at a 
somewhat higher temperature than along the direction perpendicular to the planes. 
\cite{54,55,56,57,58} The same conclusion can be gained from results of magnetization 
measurements in magnetic fields of several Oersteds. \cite{47,59,60} This justifies 
the introduction of two critical temperatures $T_c^{(ab)}$ and $T_c^{(c)}$. Below  
$T_c^{(ab)}$, the superconducting phase coherence is established along the $ab$-planes, 
but only below $T_c^{(c)}$ supercurrents can propagate in the direction of the $c$-axis. 
In the temperature range $T_c^{(c)}<T<T_c^{(ab)}$, although superconductivity already 
exists in the $ab$-planes, no mixed state can be created in magnetic fields parallel to 
the planes. This occurs only at $T\le T_c^{(c)}$, i.e., in this case our evaluation of 
$T_{c}$ corresponds to $T_c^{(c)}$. If the magnetic field is parallel to the $c$-axis, 
the mixed state can be created already at $T_c^{(ab)}$. This simple picture gives a 
natural explanation for the difference in $T_c$ for the different orientations of the 
magnetic field that was obtained for the sample Bi\#6 (see Fig. 7). The observed 
difference in $T_{c}$ is quite small ($\Delta T_{c}/T_{c} \approx 0.01$) and may easily 
be masked, for instance, by grain misalignments in grain-aligned samples. This could be 
the reason why we do not see this effect in the sample Hg\#4. It is also possible that 
this difference in $T_{c}$ is of significant magnitude only in Bi-based cuprates due to 
their very special crystalline structure. 

\section {Discussion}

The scaling procedure based on Eq. (4) turns out to be rather successful for the 
analysis of the reversible magnetization of HTSC\textquoteright s. Figs. 1 and 6 
demonstrate very well the scaling of isothermal magnetization data resulting in plots 
of the magnetization at a chosen temperature versus a renormalized magnetic field. The 
quality of scaling is remarkable for all cases that are listed in Tables I-IV and the 
mismatch between the $M(H)$ curves measured at different temperatures does not exceed 
the scatter of the original experimental data.

The most surprising result of our analysis is that for practically all families of 
HTSC\textquoteright s the $h_{c2}(T/T_{c})$ curves are virtually identical (Fig. 4). 
It is difficult to imagine that this universality of the $h_{c2}(T/T_{c})$ dependence 
is just a coincidence. We are of the opinion that the spectacular agreement between the 
$h_{c2}(T/T_{c})$ data for a great variety of different samples is an unambiguous 
evidence that our approach captures the essential features of the magnetization process 
of HTSC\textquoteright s. It does not necessarily mean, of course, that the 
Ginzburg-Landau parameter $\kappa$ is indeed temperature independent. The universality 
of $h_{c2}(T/T_{c})$ is preserved if the temperature dependence of $\kappa$ is the same 
for the different HTSC compounds studied here. 

Our analysis is applicable only to reversible magnetization data and therefore, all the 
results and conclusions are limited to temperatures close to $T_{c}$. The lower limit of 
validity, $T_{\min}$, is quite different for different families of HTSC\textquoteright s, 
as may be seen in Fig. 4. The ratio $T_{\min}/T_{c}$, which depends on the strength of 
the pinning of vortices, is highest for the Y-based compounds that exhibit the 
strongest pinning forces.

The universality of the normalized temperature dependence of $H_{c2}$ implies that the 
normalized temperature variations of the thermodynamic critical field, $H_{c}$ for 
different HTSC\textquoteright s are also identical. Since $H_c^2/8\pi$ is the difference 
in the free energy densities between the normal and superconducting states, $H_{c}(T)$ 
also reflects the temperature dependence of the superconducting energy gap $\Delta$. 
\cite{6} In other words, our result that the normalized temperature dependence of 
$H_{c2}$ follows the same universal curve for different families of HTSC\textquoteright s 
implies that the normalized temperature variations of the energy gap 
$\Delta (T/T_{c})/\Delta (0)$ for different HTSC\textquoteright s are also identical, at 
least in the temperature ranges covered in this study.

We note that the temperature dependencies of $H_{c2}$ for HTSC\textquoteright s obtained 
as a result of our analysis are qualitatively very similar to those of conventional 
superconductors. They are linear at temperatures close to $T_{c}$ with a pronounced 
negative curvature at lower temperatures. Apparently, the positive curvature of 
$H_{c2}(T)$ for HTSC\textquoteright s, which is often reported in the literature, is due 
to the uncertainty of defining $H_{c2}$ in those studies. 

\section {Conclusion}

We have developed a scaling procedure that allows to obtain the temperature 
dependence of the upper critical field from the measurements of the reversible 
isothermal magnetization. If the magnetization measurements are extended up to 
temperatures close to the superconducting critical temperature, our procedure also allows 
for a fairly reliable evaluation of the zero-field critical temperature. We have applied 
this scaling procedure for the analysis of experimental data for high-$T$ superconductors 
available in the literature and have shown that the normalized temperature dependencies 
of $H_{c2}$ are qualitatively the same as those of conventional superconductors and we 
obtain the same universal curve for different families of HTSC\textquoteright s. This 
universality is a very strong indication that also the temperature dependence of the 
superconducting energy gap is the same for all cuprate superconductors. All these 
statements have been verified to be valid at all temperatures for which data of 
measurements of the reversible magnetization of different types of 
cuprate superconductors are available in the literature.

\begin{table}[h]
\caption{Sample identification of Y-based materials.}
\begin{tabular}{lcccccc}
\colrule
\multicolumn{1}{c}{No.} &
\multicolumn{1}{c}{Refs.} &
\multicolumn{1}{c}{Compound} &
\multicolumn{1}{c}{Sample} &
\multicolumn{1}{c}{$T_{c}$ (K)} \\
\colrule
Y\#1 & 46 & YBa$_{2}$Cu$_{3}$O$_{7-x}$ & single crystal & 91.1 \\
Y\#2 & 61 & YBa$_{2}$Cu$_{3}$O$_{7}$ & single crystal & 88.0 \\
Y\#3 & 17 & YBa$_{2}$Cu$_{3}$O$_{7-x}$ & grain-aligned & 92.0 \\
Y\#4 & 16 & YBa$_{2}$Cu$_{4}$O$_{8}$ & grain-aligned & 79.8 \\
Y\#5 & 62 & YBa$_{2}$Cu$_{3}$O$_{7-x}$ & single crystal & 93.0 \\
Y\#6 & 63 & (YCa)Pb$_{2}$Sr$_{2}$Cu$_{3}$O$_{8+x}$ & single crystal & 76.0 \\
Y\#7 & 11 & YBa$_{2}$Cu$_{3}$O$_{6.94}$ & grain-aligned & 92.9 \\
Y\#8 & 11 & YBa$_{2}$Cu$_{3}$O$_{7}$ & grain-aligned & 88.7 \\
Y\#9 & 11 & YBa$_{2}$Cu$_{3}$O$_{6.85}$ & grain-aligned & 79.9 \\
Y\#10 & 5 & YBa$_{2}$Cu$_{3}$O$_{6.65}$ & single crystal & 62.3 \\
Y\#11 & 53 & YBa$_{2}$Cu$_{3}$O$_{6.5}$ & single crystal & 44.8 \\
\colrule
\end{tabular}
\end{table}
\begin{table}[h]
\caption{Sample identification of Bi-based materials.}
\begin{tabular}{lcccccc}
\colrule
\multicolumn{1}{c}{No.} &
\multicolumn{1}{c}{Refs.} &
\multicolumn{1}{c}{Compound} &
\multicolumn{1}{c}{Sample} &
\multicolumn{1}{c}{$T_{c}$ (K)} \\
\colrule
Bi\#1 & 64 & Bi$_{2}$Sr$_{2}$Ca$_{2}$Cu$_{2}$O$_{8+x}$ & single crystal &  84.0 \\
Bi\#2 & 65 & (BiPb)$_{2}$Sr$_{2}$Ca$_{2}$Cu$_{3}$O$_{x}$ & whisker & 108.7 \\
Bi\#3 & 26 & (BiPb)$_{2}$Sr$_{2}$Ca$_{2}$Cu$_{2}$O$_{8}$ & single crystal &  91.4 \\
Bi\#4 & 48 & Bi$_{2}$Sr$_{2}$Ca$_{2}$Cu$_{2}$O$_{8}$ & single crystal &  88.2 \\
Bi\#5 & 66 & Bi$_{2.1}$Sr$_{1.7}$Ca$_{1.2}$Cu$_{2}$O$_{x}$ & single crystal &  86.7 \\
Bi\#6 &  8 & Bi$_{2}$Sr$_{2}$Ca$_{2}$Cu$_{2}$O$_{8}$ & single crystal &  80.5 \\
Bi\#6 &  8 & Bi$_{2}$Sr$_{2}$Ca$_{2}$Cu$_{2}$O$_{8}$ & single crystal &  79.1 \\
\colrule
\end{tabular}
\end{table}
\newpage
\begin{table}[h]
\caption{Sample identification of Hg-based materials.}
\begin{tabular}{lcccccc}
\colrule
\multicolumn{1}{c}{No.} &
\multicolumn{1}{c}{Refs.} &
\multicolumn{1}{c}{Compound} &
\multicolumn{1}{c}{Sample} &
\multicolumn{1}{c}{$T_{c}$ (K)} \\
\colrule
Hg\# 1 & 67 & HgBa$_2$CaCu$_2$O$_{6+x}$ & grain-aligned & 117.1 \\
Hg\# 2 & 68 & HgBa$_2$Ca$_2$Cu$_4$O$_{10+x}$ & grain-aligned & 123.1 \\
Hg\# 3 & 24 & Hg$_{0.7}$Pb$_{0.3}$Sr$_2$Ca$_2$Cu$_3$O$_x$ & grain-aligned & 125.5 \\
Hg\# 5 & 13 & HgBa$_2$Ca$_2$Cu$_3$O$_{8+x}$ & grain-aligned & 131.5 \\
Hg\# 5 & 13 & HgBa$_2$Ca$_2$Cu$_3$O8+x & grain-aligned & 131.5 \\
Hg\# 6 & 21 & Hg$_{1-y}$Pb$_{y}$Ba$_{2-z}$Sr$_{z}$Ca$_2$Cu$_3$O$_x$ 
       & grain-aligned & 124.6 \\
Hg\# 7 & 71 & (HgCu)Ba$_2$CuO$_{4+x}$ & single crystal &  97.4 \\
\colrule
\end{tabular}
\end{table}
\begin{table}[h]
\caption{Sample identification of Tl-based materials.}
\begin{tabular}{lcccccc}
\colrule
\multicolumn{1}{c}{No.} &
\multicolumn{1}{c}{Refs.} &
\multicolumn{1}{c}{Compound} &
\multicolumn{1}{c}{Sample} &
\multicolumn{1}{c}{$T_{c}$ (K)} \\
\colrule
Tl\#1 & 15 & Tl$_2$Ba$_2$Ca$_2$Cu$_3$O$_{10}$ & grain-aligned & 114.6 \\
Tl\#2 & 69 & Tl$_2$Ba$_2$Ca$_2$Cu$_3$O$_{10+x}$ & grain-aligned & 122.8 \\
Tl\#3 & 70 & Tl$_{0.5}$Pb$_{0.5}$Sr$_2$CaCu$_2$O$_7$ & single crystal & 76.9 \\
Tl\#4 & 25 & TlBa$_2$Ca$_3$Cu$_4$O$_{11+x}$ & single crystal & 121.2 \\
\colrule
\end{tabular}
\end{table}


\begin{thebibliography}{71}
\bibitem{50} This function has been chosen only because it fits the experimental 
$M(H)$ data quite well and it may be replaced by any other suitable function.
\bibitem{51} The measurements at different temperatures usually cover different ranges 
of the normalized magnetic field $H/h_{c2}$. This is why it is important that only 
those data points at $T = T_{i}$, for which the values of $H/h_{c2}$ are inside the 
magnetic field range covered at $T = T_{0}$, are considered  in the fitting procedure.
\bibitem{52} The error margins indicated in this paper are solely due to the 
approximation. Taking into account experimental errors may enhance the error margins 
considerably.
\end{thebibliography}
\end{document}